\documentclass[aps,prb,groupedaddress,reprint]{revtex4-1}

\usepackage[T1]{fontenc}
\usepackage[latin9]{inputenc}

\usepackage{amsfonts,amstext,amssymb,amsmath}
\usepackage{braket}

\usepackage{graphicx,subcaption}

\usepackage{bm}

\usepackage{hyperref}

\newcommand{\B}{\Gamma}
\newcommand{\en}{\varepsilon}
\newcommand{\s}{\sigma}

\newcommand{\m}{\gamma}
\newcommand{\mr}{\tilde{\m}}

\newcommand{\NY}{\mathcal{N}}
\newcommand{\NZ}{\mathcal{M}}

\setcitestyle{numbers,square}

\begin{document}
\title{Infinite coherence time of edge spins in finite-length chains}
\author{Ivo A. Maceira$^{1}$, Fr\'ed\'eric Mila$^{1}$}
\affiliation{$^{1}$Institute of Physics, Ecole Polytechnique F\'ed\'erale de Lausanne (EPFL), CH-1015, Lausanne, Switzerland
}

\begin{abstract}
Motivated by the recent observation that exponentially long coherence times can be achieved for edge spins in models with strong zero modes, we study the impact of level crossings in finite-length spin chains on the dynamics of the edge spins. Focussing on the XY spin-$1/2$ chain with transverse or longitudinal magnetic field, two models relevant to understand recent experimental results on cobalt adatoms, we show that the edge spins can remain coherent for an infinite time even for a finite-length chain if the magnetic field is tuned to a value at which there is a level crossing. Furthermore, we show that the edge spins remain coherent for any initial state for the integrable case of transverse field because all states have level crossings at the same value of the field, while the coherence time is increasingly large for lower temperatures in the case of longitudinal field, which is non-integrable.
\end{abstract}

\maketitle

In recent experiments on chains of cobalt adatoms\citep{Toskovic2016}, level crossings of the two lowest energy states have been observed as a function of the external magnetic field $\B$. An analysis of the effective spin model of that system, the spin-1/2 XY chain with in-plane magnetic field, has revealed the presence of $N$ level crossings as a function of the magnetic field $\B$ between the two lowest energy states\citep{Dmitriev2002,Gregoire2017}. In Ref.\citep{Gregoire2017}, it was shown in particular that the model can be approximately mapped through a self-consistent mean-field method to a well-known fermionic non-interacting model, the Kitaev chain\citep{Kitaev2001}, which can in turn be described as a system of Majorana fermions coupled in pairs, with an exponentially small coupling between the two Majorana fermions located at opposite edges of the chain. For $N$ values of the magnetic field, the coupling between the two edge Majorana fermions vanishes. The two edge Majoranas can then be combined to form a zero energy regular fermion, implying that all many-particle states are degenerate. This explains in particular the ground state crossings in the spin model\citep{Gregoire2017}. In topological superconducting systems, uncoupled edge Majoranas are commonly referred to as Majorana zero modes \citep{Sarma2015}. After a Jordan-Wigner transformation the Kitaev chain becomes the XY chain with transverse field. This model has been extensively studied and its spin correlation functions\citep{Barouch1970,Barouch1971_1,Barouch1971_2,Barouch1971_3} and free energy\citep{Katsura1962} were calculated a long time ago, but the fact that there are zero modes at non-zero values of the field has only been noted recently.
 
In another recent work\citep{Fendley2017}, it was shown that "strong zero modes" associated to an ordered phase of integrable models such as the transverse field Ising (TFI) model\citep{Pfeuty1970} or the anisotropic Heisenberg XYZ model lead to a high coherence of the edge spin for long times, even for infinite temperature. The strong zero modes are operators localized at the edges of the chain that guarantee a quasi-degeneracy of all eigenstates, with a splitting that becomes exponentially small upon increasing the system size, leading to an infinite coherence time in the thermodynamic limit. A strong zero mode is still a Majorana zero mode in the sense of \citep{Sarma2015}, but the definition of strong zero mode stresses the existence of a $\mathbb{Z}_2$ symmetry which anti-commutes with the mode operator. The strong zero mode of TFI is exactly the Majorana edge quasi-particle that is decoupled from the Hamiltonian in the thermodynamic limit. When considering a perturbation that breaks the integrability of TFI, a strong zero mode could no longer be obtained, but applying the iterative method used to obtain the XYZ strong zero mode to this model resulted in an "almost strong zero mode", whose existence implies a plateau of coherence for long albeit always finite times that was observed numerically\citep{Fendley2017}. One of the perturbation terms considered was precisely a spin-spin coupling along the field, resulting in the XY chain with in-plane magnetic field. 

In this paper we explore the following idea: since degeneracies due to strong zero modes lead to a high coherence of edge spins 
that is maintained forever in the thermodynamic limit because the degeneracies become exact in that limit, then we can expect to get the same result if there are exact degeneracies for finite sizes, like in the XY model with in-plane or transverse field.

This paper is organized as follows: in Sec.~\ref{Sec:Models}, we introduce the two models we focus on, we review the exact solution of the non-interacting model and the relevant edge operators of both models, and we investigate the evolution of the level crossings as we interpolate from one model to the other. In Sec.\ref{Sec:Time_corr}, we show how the edge spin time correlation can be approximated by a single exponential (or cosine) in the ordered phase for any eigenstate, and we explore the consequences of the degeneracies for both models. We point out that the zero modes only have significant consequences for the edge spin, and we illustrate the difference numerically by comparing the correlation of edge and bulk spins. We also compare the spin time correlation of the two models for infinite temperature, where significant differences show up because the models differ by an interaction term in the fermionic language that destroys integrability.

\section{Models}
\label{Sec:Models}

Let us start by introducing the anisotropic spin-1/2 Heisenberg chain with open boundary conditions and a magnetic field $\B$ along $z$:
\begin{equation} \label{eq:H_a}
H = \sum^{N-1}_{i=1}\left( J_x \s^x_i \s^x_{i+1} + J_y \s^y_i \s^y_{i+1} + J_z \s^z_i \s^z_{i+1} \right) + \B\sum^N_{i=1} \s^z_i,
\end{equation}
where $\s^a$, $a = {x,y,z}$, are the Pauli matrices. We denote this model as XYZ-Z, with the convention that the letters before the hyphen indicate the non-zero components of the $J$ couplings, while the letter after the hyphen (if any) refers to the direction of the magnetic field if there is one. In what follows, we mostly focus on two limits of this model: XZ-Z and XY-Z, which are equivalent to an XY chain with in plane or out of plane magnetic field. Fixing the field direction and changing the couplings will prove to be more convenient when comparing the crossings of both models.

All the terms of the Hamiltonian either flip two adjacent spins or none when applied to a state with spins quantized along $z$, implying that there are no couplings between states of different $z$ spin parity. This can be quantified by the operator $P = \prod^{N}_{i=1} \s^z_i$ with eigenvalues $\pm 1$ and $[H,P] = 0$. Both models, XY-Z and XZ-Z, have an ordered phase in which the ground state is two-fold degenerate in the thermodynamic limit. For XY-Z, this phase is defined by $|\B| \leq |J_x + J_y|$. For XZ-Z, $|\B| \lesssim |J_x + J_z|$ is a good approximation for large $J_z$, while $|\B| \lesssim |J_x + \frac{3}{2} J_z|$ is more accurate for small $J_z$ \citep{Rujan1981}. For finite size, there is an energy splitting between the two lowest energy states, which belong to different $P$ parity sectors. This splitting is exponentially small with the size of the system.

\subsection{XY-Z Majorana edge fermions}

\begin{figure}[t] 
	\begin{subfigure}{0.45\textwidth}
		\includegraphics[width=\textwidth]{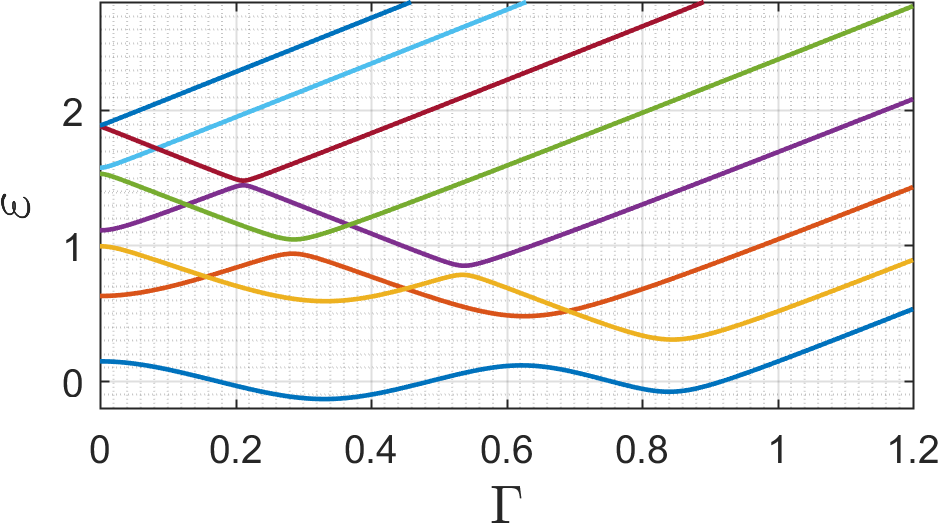} 
	\end{subfigure}
\caption{Energies $\en$ of the quasi-particle excitations of XY-Z, for $J_x= 0.6$, $J_y = 0.4$, and $N = 8$. The critical point is at $\B = J_x + J_y = 1$. The other four $\en = 0$ points are at negative $\B$, symmetric to the ones shown here.} 
\label{fig:E_XY-Z}
\end{figure} 

We review here the exact solution of XY-Z. After a Jordan-Wigner transformation into Majorana fermions and a subsequent orthogonal transformation\citep{Kitaev2001,Gregoire2017},
\begin{equation} \label{eq:s_to_maj}
		\m^a_i = \s^a_i \prod^{i-1}_{j=1} \s^z_j,\quad \s^z_i = i\m^x_i\m^y_i, \quad \mr^a_i = \sum_j Q^a_{ij}\m^a_j,
\end{equation}
where $Q^a$, $a = x,y$, are orthogonal matrices, and $\m^a, \mr^a$ obey $\{ \m^a_i, \m^b_j \} = 2\delta_{ab}\delta_{ij},~\m^a_i = (\m^a_i)^{\dagger},~ (\m^a_i)^{2} = 1$, the XY-Z model becomes a model of Majorana fermions coupled in pairs, or, equivalently, a system of free fermions with particle-hole symmetry:
\begin{equation}
H = \frac{i}{2}\sum^N_{i=1} \en_i \mr^x_i\mr^y_i = \frac{1}{2}
 \sum^N_{i=1} \en_i (\mu^{\dagger}_i\mu_i - \mu_i\mu^{\dagger}_i),
\end{equation}
where $\mu^{\dagger}_i = (\mr^x_i - i\mr^y_i)/2$ obeys the usual fermionic commutation relations. The $\en$-spectrum is illustrated in Fig.~\ref{fig:E_XY-Z}. We observe in particular that there are $\en = 0$ solutions at the fields
\begin{equation} \label{eq:B_n}
\B_n = 2\sqrt{J_x J_y}\cos\left(\frac{n\pi}{N+1}\right),
\end{equation}
with $n = 1,2,\dots N$. Note that these points only exist if $J_xJ_y > 0$. The Majorana operators $\mr^a_i$ corresponding to $\en \approx 0$, which we denote as $\mr^a$ with energy $\en_0$, are
\begin{align}	\label{eq:zero_mode_XY-Z}
	\mr^a &\approx \frac{\NY}{\lambda_+^a - \lambda_-^a}\sum^N_{n=1}((\lambda^a_+)^n - (\lambda^a_-)^{n})\m_n^a,\\
	\lambda^a_{\pm} &= \frac{-\B \pm\sqrt{\B^2-4 J_x J_y}}{2J_a},	
\end{align}
with $a = x,y$. They are exact when the energy is exactly zero, so at the points given by \eqref{eq:B_n}. The operators $\mr^x$ and $\mr^y$ are localized at the edges of the system, and one is the reflection of the other with respect to the middle of the chain. When $|J_x|>|J_y|$ (resp. $|J_y|>|J_x|$) , $\mr^x$ (resp. $\mr^y$) is localized at the first site. While we have two uncoupled Majorana fermions at the $\B_n$ points, that only means we have one zero energy fermion, resulting in the 2-fold degeneracy of all eigenstates. The Majorana edge fermions of the Ising chain (X) and TFI (X-Z) can be obtained from \eqref{eq:zero_mode_XY-Z} using appropriate limits.

Equation~\eqref{eq:B_n} guarantees the existence of a zero mode in the thermodynamic limit in the region $\B^2<4J_xJ_y$, but the full ordered phase goes beyond that. $\mr^x$ is a solution in the thermodynamic limit as long as $(\mr^x)^2 = 1$ for some $\NY$. Using this condition to calculate $\NY^2$ for $N \rightarrow \infty$ we obtained
\begin{align} \label{eq:edge_pairing_XY-Z_N}
\NY^2 &= \frac{\left(J_x-J_y\right) \left(-\B +J_x+J_y\right) \left(\B +J_x+J_y\right)}{J_x^2 \left(J_x+J_y\right)} \\
&= 1-\left(\frac{\B}{J_x}\right)^2-\left(\frac{J_y}{J_x}\right)^2 +O\left(\frac{1}{J_x^3}\right). \label{eq:edge_pairing_XY-Z_N_approx}
\end{align}
The critical lines of the $x$-ordered phase can be deduced from the condition $\NY^2 = 0$. They are given by $|\B| = |J_x+J_y|$, corresponding to the order-disorder transition, and $J_x = J_y$, the transition into the gapless XY phase. Beyond this line ($J_x < J_y$) the norm of $\mr^x$ diverges and the well-defined edge Majorana is $\mr^y$. The phase diagram was first obtained from the spin-spin correlations in  Ref.\citep{Barouch1971_1}. For a recent review of the model see Ref. \citep{Dutta2015}.

Denoting by $\ket{E}$ an eigenstate of energy $E$, we have 
\begin{equation}
\mr^x\ket{E} = (\mu^\dagger + \mu)\ket{E} = \ket{E'}, \quad \ket{E} = \mr^x\ket{E'},
\end{equation}
where $\ket{E'}$ is the eigenstate of energy $E'=E\pm\en_0$ differing from $\ket{E}$ by a quasi-particle. Each term of $\mr^x$ flips one spin when the quantization axis is along $z$, so the $P$ parity is changed. Separating the eigenstates in parity sectors, we can write
\begin{equation} \label{eq:XY-Z_pairing}
\mr^x\ket{E^{\pm}_n} = \ket{E^{\mp}_n},
\end{equation}
where $\ket{E^{\pm}_n}$ is an eigenstate with $P\ket{E^{\pm}_n} = \pm\ket{E^{\pm}_n}$.

\subsection{XZ-Z prethermal strong zero mode}
\label{sec:XZ-Z_PSZM}

The $J_z$ term of XZ-Z becomes a four fermion term after the Jordan-Wigner transformation in Eq.\ref{eq:s_to_maj}, so we no longer have a free fermion solution. In fact, the model is non-integrable, an important piece of information since integrability is believed to be a condition for the existence of a "strong zero mode"\citep{Fendley2016}. A strong zero mode ($\Psi$) is an operator that squares to 1, obeys $[H,\Psi] \sim e^{-|\alpha|N}$ and changes the $P$ parity of a state of well-defined parity. For the XY-Z model, the operator $\mr^x$ with $N \rightarrow \infty$ matches exactly this definition. In the thermodynamic limit, a strong zero mode commutes with the Hamiltonian but changes the parity of the state. So each level must contain a state of each symmetry, and the spectrum of both sectors are identical. This is the case of XYZ, which has a strong zero mode inside the ordered phase\citep{Fendley2016}.

The XZ-Z model does not have a strong zero mode, but it has an "almost strong zero mode"\citep{Fendley2017}, later understood as a "prethermal strong zero mode"\citep{Else2017}, implying the emergence of a conserved quantity for a quasi-exponential time\citep{Abanin2017_1,Abanin2017_2}. Such an operator, which we denote as $\Phi$, has the same properties as a strong zero mode except that the commutator is always finite: $[H,\Phi] = \nu$, where $\nu$ is an operator whose norm decreases exponentially with the size up to some limiting system size where a minimum is reached. Using this commutator we have
\begin{equation}
(H\Phi - \nu) \ket{E^{\pm}_n} = E^{\pm}_n \Phi \ket{E^{\pm}_n} 
\end{equation}
for an eigenstate $\ket{E^{\pm}_n}$. Assuming that the norm of $\nu$ is sufficiently small, we may write
\begin{equation}
\Phi\ket{E^{\pm}_n} \approx \ket{E^{\mp}_n},
\end{equation}
with $E^{\pm}_n-E^{\mp}_n \sim \lVert \nu \rVert$. In the limit $J_z = 0$, $\Phi$ would become the X-Z edge Majorana fermion $\mr^x$ and we would recover Eq.~\ref{eq:XY-Z_pairing}. The operators $\Phi$ and $\mr^x$ have an important similarity in that their leading operator is the same:
\begin{align}
\mr^x &= \NY \s^x_1 + \dots, \quad \Phi = \NZ \s^x_1 + \dots,\\
\NZ^2 &= 1-\left(\frac{\B}{J_x}\right)^2-\left(\frac{J_z}{J_x}\right)^2 +O\left(\frac{1}{J_x^3}\right).
\end{align}
The second order expansions of their normalization constants $\NY$ and $\NZ$ are also identical. In the limit of the Ising model ($J_y=J_z=\B=0$) both operators become equal to $\s^x_1 = \m^x_1$, which is exactly the uncoupled edge Majorana fermion of that model. As we will see, the existence of the operators $\mr^x$ and $\Phi$ together with the level crossings are the factors that allow a high coherence of the edge spins for an infinite time for both models. 

\subsection{Level crossings}

\begin{figure*}[t]
	\begin{subfigure}{0\textwidth}
		\phantomsubcaption{~}
		\label{fig:E_crossings_1}
	\end{subfigure}
	\begin{subfigure}{0\textwidth}
		\phantomsubcaption{~}
		\label{fig:E_crossings_2}
	\end{subfigure}
	\begin{subfigure}{0\textwidth}
		\phantomsubcaption{~}
		\label{fig:E_crossings_3}
	\end{subfigure}
	\begin{subfigure}{0\textwidth}
		\phantomsubcaption{~}
		\label{fig:E_crossings_4}
	\end{subfigure}
	\begin{subfigure}{0\textwidth}
		\phantomsubcaption{~}
		\label{fig:E_crossings_5}
	\end{subfigure}
	\begin{subfigure}{\textwidth}
	\includegraphics[width=\textwidth]{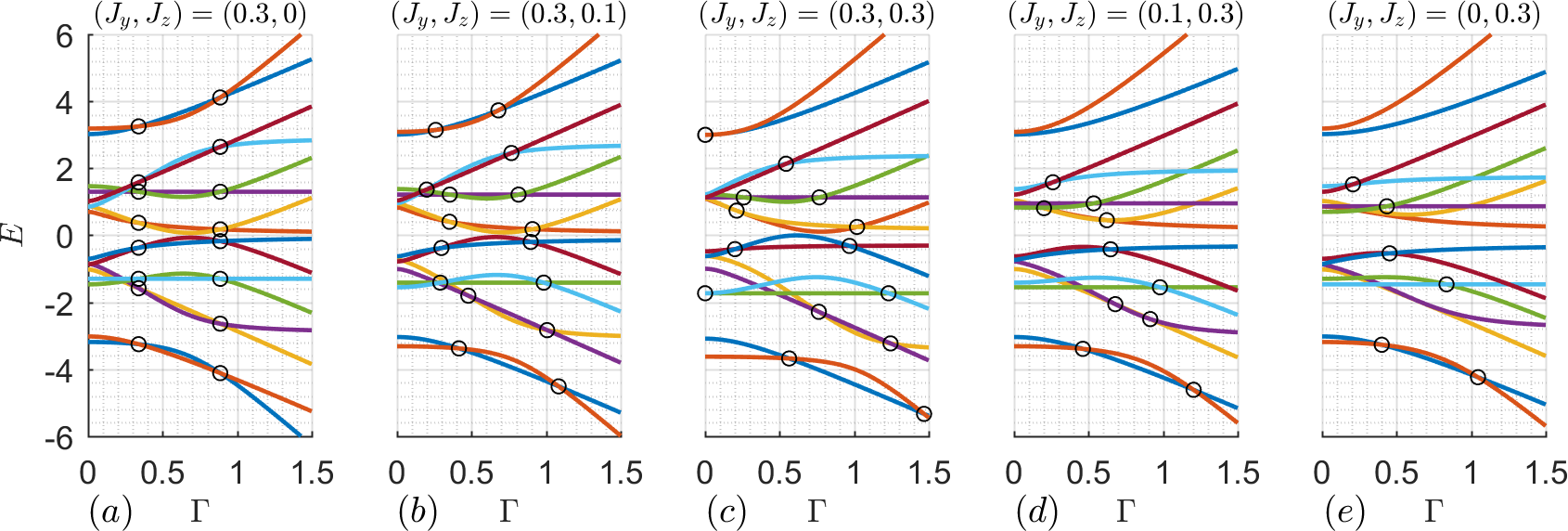}
	\end{subfigure}
\caption{Spectrum of XYZ-Z for $N = 4$, $J_x=1$, and several values of ($J_y,J_z$) with XY-Z on the left and XZ-Z on the right.}
\label{fig:E_crossings}
\end{figure*}

The addition of a $J$ coupling to X-Z, be it $J_y$ or $J_z$, creates oscillations in the energies as a function of $\B$ inside the ordered phase, which causes crossings between pairs of quasi-degenerate states of different parity, a behaviour not present in the TFI. In particular, both models have $N$ points of exact ground state degeneracy as a function of $\B$ in some parameter region. The ground state crossings of XZ-Z have already been studied in detail\citep{Gregoire2017}: when $|J_x| > |J_z|$, the two lowest energy states form a low energy sector isolated from the rest, and if $J_z > 0$, there are $N$ crossings between these two states, while there are no crossings for $J_z < 0$. However, the spectrum of $H_z$ is independent of the sign of $J_x$ and $\B$, so the spectrum of $-H(J_z)$ is the same as $H(-J_z)$, implying that for $J_z < 0$ the crossings are present in the highest energy state.

Depending on which $J$ coupling is the largest and on its sign, we have different ordered phases. We study the phase $|J_x| > |J_{y,z}|$ where there is order in $x$. The signs of the couplings are not very important for the correlation, but we want crossings to exist in the ground state. So, from now on we restrict ourselves to $J_x > J_{y,z} > 0$. Also, the physical situations of positive or negative field $\B$ are equivalent by rotation, so we only discuss $\B > 0$.

In Fig.~\ref{fig:E_crossings} we show how the crossings in each model are adiabatically related to each other: starting from the XY-Z model (Fig.~\ref{fig:E_crossings_1}), where the non-interacting nature is noticeable and where all energies are degenerate at $\B_n$, and adding a coupling in the $z$ direction that obeys $J_x > J_z > 0$, the ground state crossings continuously move towards higher $\B$. Then when decreasing $J_y \rightarrow 0$ they become the $N$ crossings of XZ-Z (Fig.~\ref{fig:E_crossings_5}). The middle spectrum (Fig.~\ref{fig:E_crossings_3}) corresponds to $J_y = J_z$, which is a turning point where some of the crossings disappear if we start from XY-Z and increase $J_z$. In particular, both crossings of the highest energy pair meet at $\B = 0$, after which a gap appears between these states. When $J_y$ finally becomes zero, a second set of crossings vanishes. The energy pairing in XZ-Z is highly asymmetrical: lower energy pairs have a small gap up to fields much higher than their high energy counterparts. Note that the roles would be reversed for negative $J_z$. Changing the sign of $J_z$ would invert the spectra in energy, and the ground state of XZ-Z would have no crossing. The three middle plots in Fig.~\ref{fig:E_crossings} show that the XYZ-Z model also has energy crossings in some parameter region, implying that some of our results could be extended to the more general case.

\section{Edge spin time correlation}
\label{Sec:Time_corr}

This section is devoted to analytical and numerical results regarding the auto-correlation of edge spins. Following Ref.\citep{Fendley2017}, we consider the edge spin time auto-correlation of an eigenstate $\ket{E_n^{\pm}}$ of energy $E_n^{\pm}$ and $P$ parity $\pm 1$ defined by
\begin{equation} \label{eq:edge_spin_corr}
		A_n^{\pm}(t) \equiv \bra{E_n^{\pm}} \s^x_1(t) \s^x_1 (0) \ket{E_n^{\pm}}
\end{equation}
where the component of the spin is that along which the system is ordered.
We introduce $I = \sum_{m} \ket{E_m^{+}}\bra{E_m^{+}} + \ket{E_m^{-}}\bra{E_m^{-}}$ to obtain
\begin{equation} \label{eq:spin_corr_closure_rel}
	\begin{split}
		A_n^{\pm}(t) &= \bra{E_n^{\pm}} e^{-iHt} \s^x_1 e^{iHt} I \s^x_1 \ket{E_n^{\pm}}\\
		 &= \sum_m |\bra{E_m^{\mp}} \s^x_1 \ket{E_n^{\pm}}|^2 e^{i(E_m^{\mp} - E_n^{\pm})t}.
	\end{split}
\end{equation}
In this form it becomes obvious that any degeneracy creates time-independent positive terms in the correlation, as long as the appropriate matrix element is non-zero. However, not only is the matrix element between the states with crossings circled in Fig. \ref{fig:E_crossings} non-zero, but we found that it dominates over all other matrix elements while inside the bulk of the ordered phases, implying that we have both degeneracies and a high value of coherence.

Coherence can still be present when considering higher temperatures\citep{Fendley2017}. The limit of lowest coherence should be at infinite temperature, where the average edge spin correlation will be
\begin{equation}
\bar{A}(t) \equiv \braket{\s^x_1(t) \s^x_1}_{T=\infty} = \frac{1}{2^N} \sum_{n} [A_n^{+}(t) + A_n^{-}(t)]. 
\end{equation}

\subsection{XY-Z correlation}

The correlation of XY-Z can be determined exactly since $\s^x_1$ is exactly the local Majorana fermion $\m^x_1$, which, inverting the last equation of \eqref{eq:s_to_maj}, is given by
\begin{equation}
\s^x_1 = \m^x_1 = \sum_k Q^x_{k1}\mr^x_k,
\end{equation} 
where we sum over the $N$ Majorana fermions, one of them being the edge Majorana $\mr^x$. Substituting in \eqref{eq:spin_corr_closure_rel}, we get
\begin{align}
	A_n^{\pm}(t) &= \sum_{m} |\bra{E_m^{\mp}} \sum_{k} Q^x_{k1}\mr^x_k \ket{E_n^{\pm}}|^2 e^{i(E_m^{\mp} - E_n^{\pm})t} =	\nonumber \\
		&= \sum_{k} |Q^x_{k1}|^2 e^{i g_{nk}^{\pm} \en_k t}
\end{align}
where $g_{nk}^{\pm} \equiv -i\bra{E_n^{\pm}}\mr^x_k\mr^y_k\ket{E_n^{\pm}}$ is equal to $-1$ or $1$ depending on whether the fermion $\mu_k^{\dagger}$ is present in the state or not. The correlation of any state consists of the same $N$ terms with different signs in the exponentials. By symmetry, the result must be the same at the other edge of the chain. Note that we cannot write the same decomposition for the correlation of spins in the bulk since only $\s^x_1$ corresponds directly to one of the local Majorana fermions in the Jordan-Wigner transformation. So we expect a difference between edge and bulk spins. 

\begin{figure}[t] 
	\begin{subfigure}[t]{0.5\columnwidth} 
	\includegraphics[width=0.95\columnwidth]{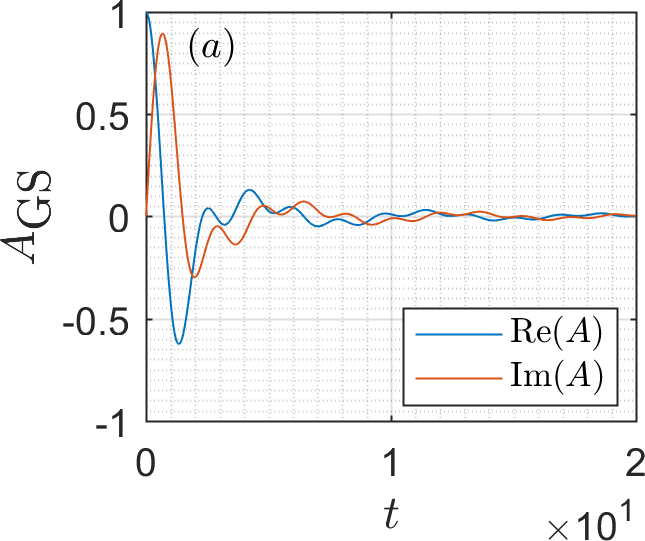}
	\phantomsubcaption{~}
	\label{fig:A_XY-Z_dis}
	\end{subfigure}%
	\begin{subfigure}[t]{0.5\columnwidth} 
	\includegraphics[width=0.95\columnwidth]{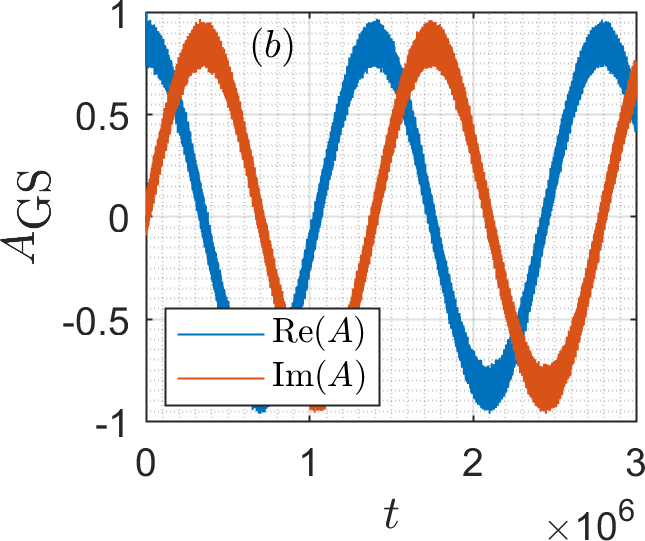}
	\phantomsubcaption{~}
	\label{fig:A_XY-Z_ord}
	\end{subfigure}
	\begin{subfigure}{0.5\columnwidth} 
	\includegraphics[width=0.95\columnwidth]{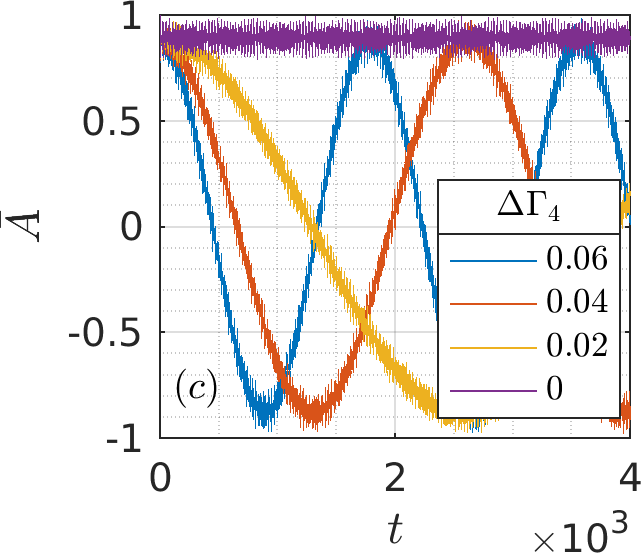}
	\phantomsubcaption{~}
	\label{fig:A_XY-Z_near_zero}
	\end{subfigure}%
	\begin{subfigure}{0.5\columnwidth} 
	\includegraphics[width=0.95\columnwidth]{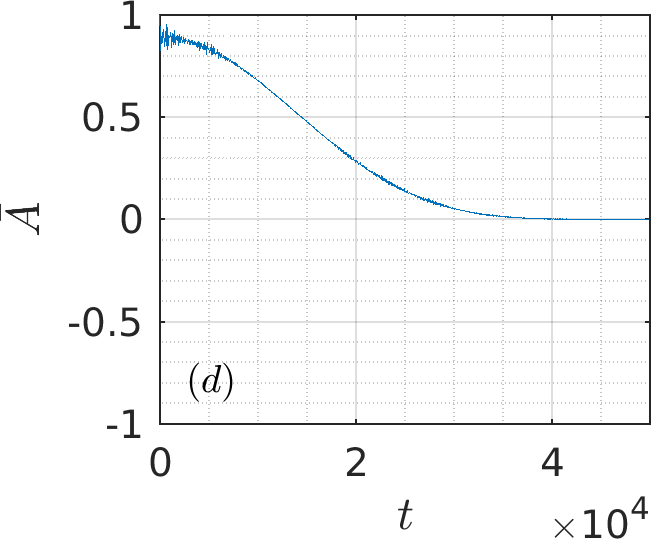}
	\phantomsubcaption{~}
	\label{fig:A_XY-Z_perturbed}
	\end{subfigure}
\caption{(a,b) Ground state edge spin correlation of XY-Z for $N = 20$, $J_x = 1$, $J_y = 0.3$, (a) in the disordered phase ($\B = 2.2$), (b) in the ordered phase ($\B = 0.35$). (c,d) Average edge spin correlation for $N = 8$, $J_x = 1$, $J_y = 0.3$, (c) close to $\B_4 \approx 0.19$, (d) for $\B_4 \approx 0.19$, but with $J_z = 0.001$.}
\label{fig:A_XY-Z}
\end{figure}

In the disordered phase, there is no Majorana fermion that is localized at the edge, so all terms are of the same order of magnitude but differ in amplitude and frequency. Accordingly, the system quickly becomes decoherent (Fig.~\ref{fig:A_XY-Z_dis}). In the thermodynamic limit, assuming that all modes have the same amplitude at the edge spin and considering the ground state correlation in which $g_{nk}^{\pm} = 1$ for all $k$, we have
\begin{equation} \label{Eq:corr_band}
	\begin{split}
A_{\text{GS}}(t) \approx \frac{1}{\en_t - \en_b} \int_{\en_b}^{\en_t} e^{i \en t} d\en = -\frac{i}{t}\frac{e^{i\en_t t} - e^{i\en_bt} }{\en_t - \en_b},
	\end{split}
\end{equation}
where $\en_t$ and $\en_b$ are the limits of the band, leading to $A_{\text{GS}}(\infty) = 0$. We expect the same result for all states. In the ordered phase, the edge mode term stands out in amplitude and frequency. Writing explicitly the $\mr^x$ term we have
\begin{equation}
A_n^{\pm}(t) = \NY^2 e^{i g_{n0}^{\pm} \en_0 t} + z(t),
\end{equation}
where $|z(t)| \leq (1-\NY^2)$ and $z(t)$ is the bulk contribution to the correlation which, as we saw in \eqref{Eq:corr_band}, disappears for infinite $N$ and $t$, so that $A_n^{\pm}(\infty) = \NY^2$ in the thermodynamic limit. For finite sizes, $\en_0$ can be orders of magnitude lower than the other energies, so the term $z(t)$ looks like noise on the time-scale of $1/\en_0$ (Fig.~\ref{fig:A_XY-Z_ord}), even though it is well-defined. We can thus approximate 
\begin{equation}
A_n^{\pm}(t) \approx \NY^2 e^{i g_{n0}^{\pm} \en_0 t}.
\end{equation}
So the edge spin flips after an interval of time $\tau = \pi/\en_0$, independently of the eigenstate the system is in. Close to the $\B_n$ points of Eq.~\eqref{eq:B_n}, $\en_0$ is approximately linear with $\Delta \B_n = \B - \B_n$, so $\tau \sim 1/\Delta\B_n $. Since $\en_0$ is exponentially suppressed with system size, we have $\tau \sim e^{|\alpha|N}/\Delta\B_n $, allowing for a better fine-tuning of the coherence time for larger sizes. Exactly at $\B_n$ we have $A_n^{\pm}(t) \approx \NY^2$, so the edge spin remains coherent for an infinite time (Fig.~\ref{fig:A_XY-Z_near_zero}). Even for infinite temperature, we have
\begin{equation}
\bar{A}(t) = \textrm{Re}(A_n^{\pm}(t)) \approx \NY^2 \cos(\en_0 t).
\end{equation}
so the same discussion applies in this limit. However, this result is very sensitive to any realistic perturbation. For example, adding a very small $J_z$ coupling does not alter significantly $\NY^2$, but each pair of states will have a slightly different energy difference and the crossings will move away from $\B_n$ as we saw in Fig.~\ref{fig:E_crossings} so that at some point we must reach decoherence, and an infinitely-lived plateau is no longer present at the $\B_n$ points (Fig.~\ref{fig:A_XY-Z_perturbed}). However, if we were to change the field slightly to a value where one of the crossings moved to, then we would recover a (small) positive constant term in the correlation and the coherence time would be infinite again. We explore this fact in more detail in the next section. 

\begin{figure}[t] 
	\begin{subfigure}{0.5\columnwidth} 
	\includegraphics[width=0.95\columnwidth]{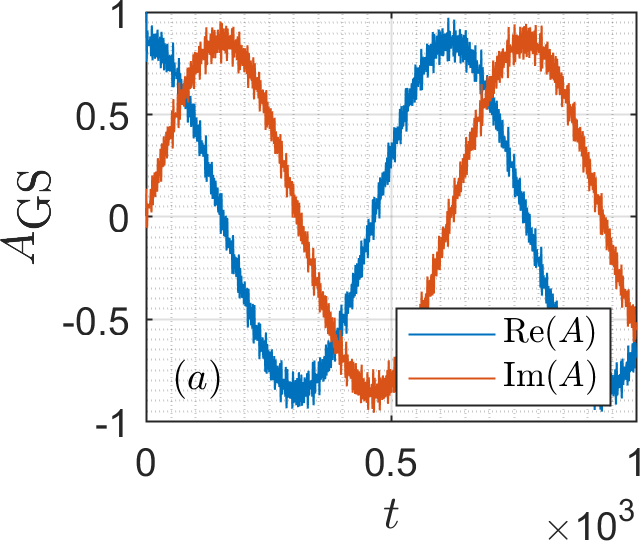}
	\phantomsubcaption{~}
	\label{fig:A_XZ-Z_ground}
	\end{subfigure}%
	\begin{subfigure}{0.5\columnwidth} 
	\includegraphics[width=0.95\columnwidth]{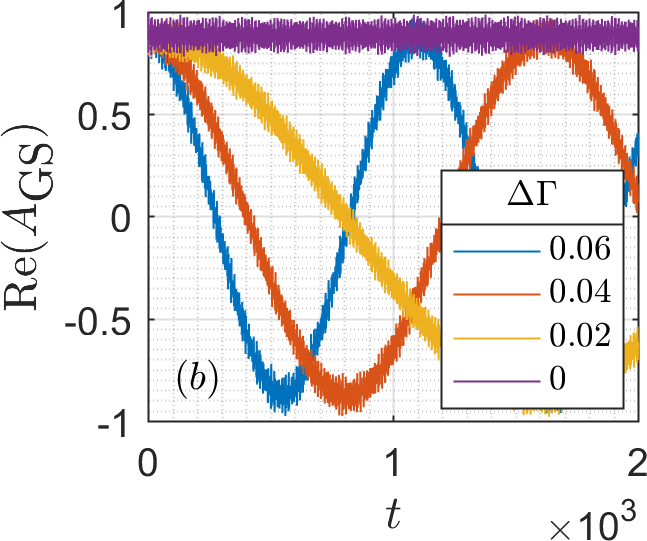}
	\phantomsubcaption{~}
	\label{fig:A_XZ-Z_near_zero}
	\end{subfigure}
	\begin{subfigure}{0.5\columnwidth} 
	\includegraphics[width=0.95\columnwidth]{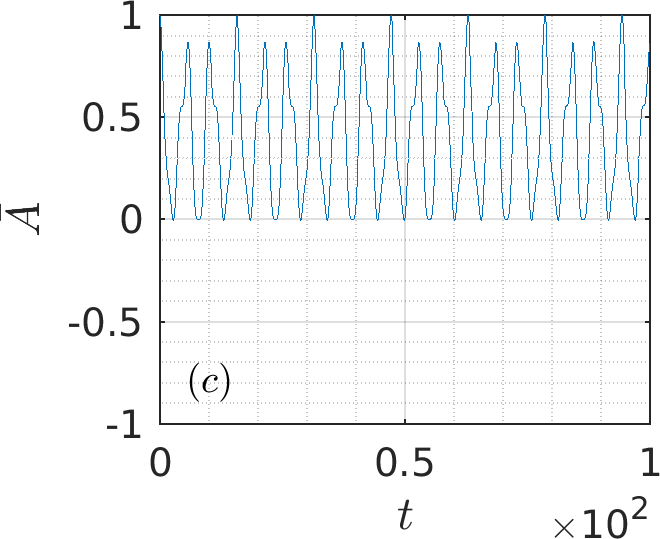}
	\phantomsubcaption{~}
	\label{fig:A_XZ-Z_crossing_2}
	\end{subfigure}%
	\begin{subfigure}{0.5\columnwidth} 
	\includegraphics[width=0.95\columnwidth]{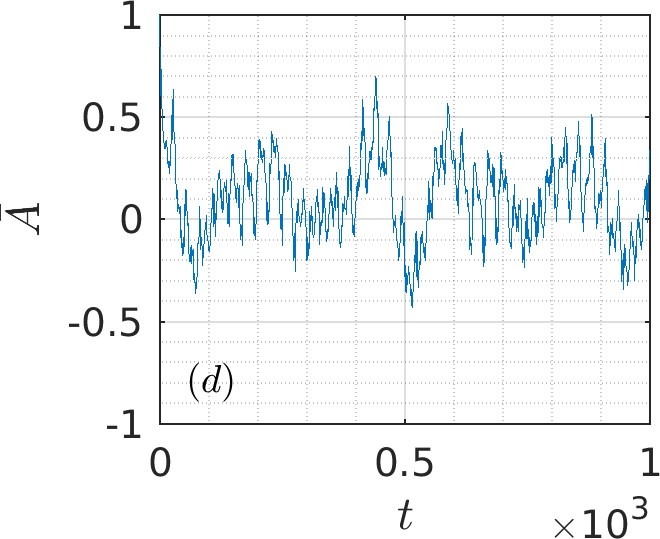}
	\phantomsubcaption{~}
	\label{fig:A_XZ-Z_crossing_4}
	\end{subfigure}
\caption{(a),(b) Ground state edge spin correlation for $N = 8$, $J_x = 1$, $J_z = 0.3$ (a) in the ordered phase ($\B = 0.35$), (b) approaching the first ground state crossing at $\B \approx 0.2341$. (c,d) Average edge spin time correlation at ground state crossings, with $J_x = 1$, $J_z = 0.3$ and (c) $N=2$, $\B \approx 0.62$, (d) $N=4$, $\B \approx 0.4$.} 
\label{fig:A_XZ-Z}
\end{figure} 

\subsection{XZ-Z correlation}

We cannot obtain any exact result for the XZ-Z correlation, but as we saw in Sec.~\ref{sec:XZ-Z_PSZM} there is an operator that gives us a pairing between states of different parity of the form $\Phi\ket{E_n^{\pm}} \approx \ket{E_n^{\mp}}$. Using this with the properties of $\s^x_1$, we have
\begin{align} \label{eq:edge_mat_elem}
&\bra{E_n^{\mp}} \Phi \s^x_1 \ket{E_n^{\mp}} \approx \bra{E_n^{\pm}} \s^x_1 \ket{E_n^{\mp}} \approx \bra{E_n^{\pm}} \s^x_1 \Phi \ket{E_n^{\pm}} \Rightarrow \nonumber \\
&\bra{E_n^{\pm}} \s^x_1 \ket{E_n^{\mp}} \approx \frac{1}{2}\bra{E_n^{\pm}} \{\s^x_1,\Phi \} \ket{E_n^{\pm}}.
\end{align} 
If $J_z = 0$, the expression would have no error term, $\Phi$ would become $\mr^x$, and the anti-commutator would be a constant: $\{\s^x_1,\mr^x \} = 2\NY$. Using the next-order terms of $\Phi$ determined in \citep{Fendley2017}, we find $\{\s^x_1,\Phi \} = 2\NZ + \hat{O}\left(J^2_z/J^2_x\right)$, and substituting in \ref{eq:edge_mat_elem} we have
\begin{equation} \label{eq:edge_mat_elem_2}
\bra{E_n^{\pm}} \s^x_1 \ket{E_n^{\mp}} = \NZ + O\left( \frac{J^2_z}{J^2_x} \right)
\end{equation}
In the ordered phase, the correlation can be approximated by its main term:
\begin{equation}
	\begin{split}
A_n^{\pm}(t) &\approx |\bra{E_n^{\mp}} \s^x_1 \ket{E_n^{\pm}}|^2 e^{i(E_n^{\mp} - E_n^{\pm})t} \\
	&\approx \NZ^2 e^{i(E_n^{\mp} - E_n^{\pm})t},
	\end{split}
\end{equation}
where in the first equation we ignore all other terms and in the second equation we use \ref{eq:edge_mat_elem_2}. This result is confirmed numerically, as seen in Fig.~\ref{fig:A_XZ-Z_ground}. 
For finite sizes, we have $A_n^{\pm}(t) \approx \NZ^2$ when the paired states are degenerate, which happens $N$ times for the ground state (Fig.~\ref{fig:A_XZ-Z_near_zero}). The discussion regarding the XY-Z coherence time close to the degeneracy points is also applicable here. However, by contrast with XY-Z, the time-independent term of the average correlation can be quite small since the crossings of different pairs do not happen for the same field. At a pair crossing we have
\begin{equation}
	\bar{A}(t) \approx \NZ^2/2^{N-1} + f(t),
\end{equation}
for some real function $|f(t)|\leq 1-\NZ^2/2^{N-1}$ whose time average is approximately zero. The constant term may not be noticeable due to the noise $f(t)$. The constant term could be doubled or, although very unlikely, tripled, if for certain J couplings there are coincident crossings. In Figs.~\ref{fig:A_XZ-Z_crossing_2} and \ref{fig:A_XZ-Z_crossing_4} we show $\bar{A}$ at the ground state crossings of very small chains. The time average in both cases gives approximately the expected constant term, but it is clear that the constant term will be harder to detect under the noise as we increase the chain size.

\subsection{Edge vs. bulk}

\begin{figure}[t] 
	\begin{subfigure}{0.48\textwidth}
		\includegraphics[width=\textwidth]{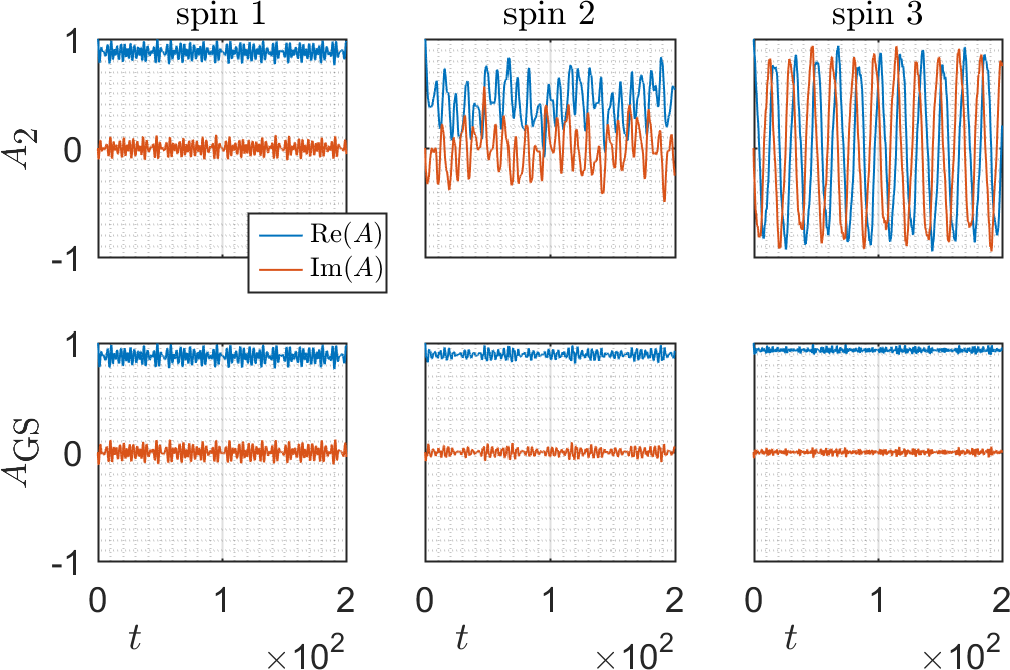}
	\end{subfigure}
\caption{Spin time correlation of the first 3 spins ($N = 6$) on the first and second ($A_2$) lowest-energy pair of states of XY-Z, with $J_x=1$, $J_y = 0.3$, at the respective crossings of lowest $\B$ ($\B \approx 0.244$).}
\label{fig:A_XY-Z_excited}
\end{figure} 

\begin{figure}[t] 
	\begin{subfigure}{0.48\textwidth}
		\includegraphics[width=\textwidth]{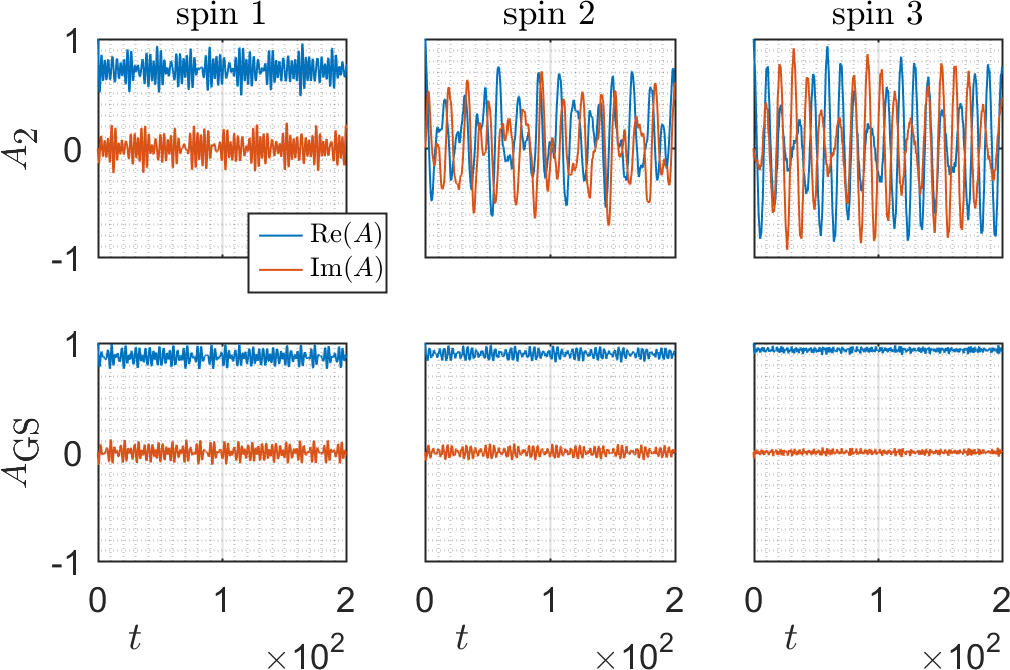}
	\end{subfigure}
\caption{Spin time correlation of the first 3 spins ($N = 6$) on the first and fifth ($A_5$) lowest-energy pair of states of XZ-Z, with $J_x=1$, $J_z = 0.3$, at their respective crossings of lowest $\B$ ($\B_{\textrm{ground}} \approx 0.295$, $\B_{\textrm{excited}} \approx 0.376$).}
\label{fig:A_XZ-Z_excited}
\end{figure}

The fact that the main term of the $\mr^x$ and $\Phi$ operators is $\sigma^x_1$ has important consequences for the edge spin, but that is the only term which is a single Pauli matrix, all others being products of Pauli matrices. So we cannot conclude anything about the bulk correlation from them. To highlight the difference between edge and bulk spins, we show in Figs.~\ref{fig:A_XY-Z_excited} (XY-Z) and \ref{fig:A_XZ-Z_excited} (XZ-Z) the correlation along the spin chain at a crossing point of the ground state and of a pair of excited states. While the ground state correlation is even higher and consequently has less noise in the bulk, this behavior is mainly lost in the excited states, but some state pairing is still manifest. For example, on the second spin of the XY-Z chain, the plateau visible for the first excited state (index $n=2$) pair implies that the term $|\bra{E_2^{+}} \s^x_2 \ket{E_2^{-}}|^2 \approx 0.5$ dominates over the rest. The third spin has no plateau, but the $\s^x_3$ elements reveal a pairing in $|\bra{E_4^{\pm}} \s^x_3 \ket{E_2^{\mp}}|^2 \approx 0.9$, resulting in a correlation that can be approximated by $0.9 e^{i(E_4^{\pm}-E_2^{\mp})t}$. 

The fact that the coherence is maintained at all sites in the ground state is easy to understand in the limit of the slightly perturbed Ising model (i.e. small  $J_y$ and $\B$). In that limit, the two quasi-degenerate ground states are given by
\begin{equation}
\ket{E_1^{\pm}} \sim \frac{1 \pm P}{\sqrt{2}}\ket{\rightarrow \leftarrow \rightarrow \leftarrow \rightarrow \leftarrow}_x,
\end{equation}
where the spins are along the $x$ direction. Calculating the matrix elements explicitly from here, and noting that $\{\s_n^x,P\} = 0$, we have $|\bra{E_1^{+}} \s^x_n \ket{E_1^{-}}|^2 \sim 1$, from which the ground state coherence of all spins follows.  
It would be interesting to see to which extent the observations for the excited states can be rationalized along similar lines. This goes beyond the scope of the present paper however and is left for future investigation.

\subsection{Summary and discussion}

In both the XY-Z and the XZ-Z models, the edge spin time correlation of any eigenstate can be simplified to a single exponential in the ordered phase due to Majorana-like operators localized at the edges that commute or almost commute with the Hamiltonian. At the crossing points of two paired states the edge spin state is partially conserved: its time autocorrelation does not decay to zero but goes to a finite value (generically smaller than 1) in the limit of infinite time. Close to the crossing point the edge spin seems to be rotating with a period that is proportional to $1/\Delta \B$. This could prove of experimental relevance since the edge spin can be controlled by an external magnetic field. 

While all the spins of the chain show coherence at the crossings in the ground state, the edge spins are different in that they are coherent in any excited state. As a consequence, the coherence remains relatively unaltered for XY-Z at high temperatures, although the system becomes more sensitive to perturbations the higher the temperature. For XZ-Z, coherence can be maintained at a plateau of value $\NZ^2$ for long times as found in \citep{Fendley2017}, after which it decays to $n\NZ^2/2^{N-1}$ if there are $n$ degenerate pairs for the current field. Thus for any temperature there has to be a plateau between $\NZ^2$ and $\NZ^2/2^{N-1}$. 

\acknowledgements FM is thankful to Paul Fendley for very instructive discussions. This work has been supported by the Swiss National Science Foundation and the Portuguese Science and Technology Foundation (FCT) through the grant SFRH/BD/117343/2016.

\bibliographystyle{apsrev4-1}
\bibliography{bib}

\end{document}